\documentstyle[12pt]{article}

\skewchar\fivmi='177
\skewchar\sixmi='177
\skewchar\sevmi='177
\skewchar\egtmi='177
\skewchar\ninmi='177
\skewchar\tenmi='177
\skewchar\elvmi='177
\skewchar\twlmi='177
\skewchar\frtnmi='177
\skewchar\svtnmi='177
\skewchar\twtymi='177
\def\@magscale#1{ scaled \magstep #1}
\font\twfvmi  = ammi10   \@magscale5 % math italic
\skewchar\twfvmi='177
\skewchar\fivsy='60
\skewchar\sixsy='60
\skewchar\sevsy='60
\skewchar\egtsy='60
\skewchar\ninsy='60
\skewchar\tensy='60
\skewchar\elvsy='60
\skewchar\twlsy='60
\skewchar\frtnsy='60
\skewchar\svtnsy='60
\skewchar\twtysy='60
\font\twfvsy  = amsy10   \@magscale5 % math symbols
\skewchar\twfvsy='60

\catcode`@=11
\def\un#1{\relax\ifmmode\@@underline#1\else
        $\@@underline{\hbox{#1}}$\relax\fi}
\catcode`@=12

\let\du=\d                      % dot-under
\let\um=\H                      % Hungarian umlaut
\def\a{\alpha}
\def\b{\beta}

\def\d{\delta}
\def\e{\epsilon}

\def\g{\gamma}

\def\k{\kappa}
\def\l{\lambda}
\def\m{\mu}

\def\s{\sigma}

\def\D{\Delta}

\def\G{\Gamma}

\def\L{\Lambda}

\font\sc=font005                        % script
\def\Sc#1{{\hbox{\sc #1}}}      % script for single characters in equations
\font\ooo=circle10                      % thin circles
\font\ro=manfnt                         % font with rope
\def\kcl{{\hbox{\ro 6}}}                % left-handed rope
\def\kcr{{\hbox{\ro 7}}}                % right-handed rope
\def\ktl{{\hbox{\ro \char'134}}}        % top end for left-handed rope
\def\ktr{{\hbox{\ro \char'135}}}        % " right
\def\kbl{{\hbox{\ro \char'136}}}        % " bottom left
\def\kbr{{\hbox{\ro \char'137}}}        % " right
\def\ip{{=\!\!\! \mid}}                                    % 2-d plus
\def\bo{{\raise.15ex\hbox{\large$\Box$}}}               % D'Alembertian
\def\pr{\prod}                                          % product
\def\TH{{\raise.2ex\hbox{$\displaystyle \bigodot$}\mskip-4.7mu \llap H \;}}
\def\face{{\raise.2ex\hbox{$\displaystyle \bigodot$}\mskip-2.2mu \llap {$\ddot
        \smile$}}}                                      % happy face
\def\sp#1{{}^{#1}}                              % superscript (unaligned)
\def\Tilde#1{\widetilde{#1}}                    % big tilde
\def\Hat#1{\widehat{#1}}                        % big hat
\def\Bar#1{\overline{#1}}                       % big bar
\def\leftrightarrowfill{$\mathsurround=0pt \mathord\leftarrow \mkern-6mu
        \cleaders\hbox{$\mkern-2mu \mathord- \mkern-2mu$}\hfill
        \mkern-6mu \mathord\rightarrow$}
\def\dvec#1{\vbox{\ialign{##\crcr
        \leftrightarrowfill\crcr\noalign{\kern-1pt\nointerlineskip}
        $\hfil\displaystyle{#1}\hfil$\crcr}}}           % <--> accent
\def\dt#1{{\buildrel {\hbox{\LARGE .}} \over {#1}}}     % dot-over for sp/sb
\def\frac#1#2{{\textstyle{#1\over\vphantom2\smash{\raise.20ex
        \hbox{$\scriptstyle{#2}$}}}}}                   % fraction
\def\ha{\frac12}                                        % 1/2
\def\sfrac#1#2{{\vphantom1\smash{\lower.5ex\hbox{\small$#1$}}\over
        \vphantom1\smash{\raise.4ex\hbox{\small$#2$}}}} % alternate fraction
\def\bfrac#1#2{{\vphantom1\smash{\lower.5ex\hbox{$#1$}}\over
        \vphantom1\smash{\raise.3ex\hbox{$#2$}}}}       % "
\def\afrac#1#2{{\vphantom1\smash{\lower.5ex\hbox{$#1$}}\over#2}}    % "
\newskip\humongous \humongous=0pt plus 1000pt minus 1000pt
\def\caja{\mathsurround=0pt}
\def\eqalign#1{\,\vcenter{\openup2\jot \caja
        \ialign{\strut \hfil$\displaystyle{##}$&$
        \displaystyle{{}##}$\hfil\crcr#1\crcr}}\,}
\newif\ifdtup
\def\panorama{\global\dtuptrue \openup2\jot \caja
        \everycr{\noalign{\ifdtup \global\dtupfalse
        \vskip-\lineskiplimit \vskip\normallineskiplimit
        \else \penalty\interdisplaylinepenalty \fi}}}
\def\li#1{\panorama \tabskip=\humongous                         % eqalignno
        \halign to\displaywidth{\hfil$\displaystyle{##}$
        \tabskip=0pt&$\displaystyle{{}##}$\hfil
        \tabskip=\humongous&\llap{$##$}\tabskip=0pt
        \crcr#1\crcr}}

\def\ref#1{$\sp{#1)}$}

\parskip=\medskipamount                 % space between paragraphs (LaTeX)
\thicklines                         % thick straight lines for pictures (LaTeX)

\def\oldheadpic{                                % old UM heading
        \setlength{\unitlength}{.4mm}
        \thinlines
        \par
        \begin{picture}(349,16)
        \put(325,16){\line(1,0){4}}
        \put(330,16){\line(1,0){4}}
        \put(340,16){\line(1,0){4}}
        \put(335,0){\line(1,0){4}}
        \put(340,0){\line(1,0){4}}
        \put(345,0){\line(1,0){4}}
        \put(329,0){\line(0,1){16}}
        \put(330,0){\line(0,1){16}}
        \put(339,0){\line(0,1){16}}
        \put(340,0){\line(0,1){16}}
        \put(344,0){\line(0,1){16}}
        \put(345,0){\line(0,1){16}}
        \put(329,16){\oval(8,32)[bl]}
        \put(330,16){\oval(8,32)[br]}
        \put(339,0){\oval(8,32)[tl]}
        \put(345,0){\oval(8,32)[tr]}
        \end{picture}
        \par
        \thicklines
        \vskip.2in}
\def\oldtitle#1#2#3#4{\oldheadpic\begin{center}\vglue.5in{\large\bf #1}\\[.6in]
        {#2}\\[.1in] {\it Department of Physics and Astronomy}\\
        {\it University of Maryland, College Park, MD 20742}\\[.6in]
        Physics Publication \#{#3}\\ {#4}\\[1.5in] {\bf ABSTRACT}\\[.1in]
        \end{center} \begin{quotation}}                 % old title stuff
\def\oldTitle#1#2#3#4#5#6#7{\oldheadpic\begin{center} \vglue .4in
        {\large\bf #1}\\[.4in]
        {#2}\\[.1in] {\it Department of Physics and Astronomy}\\
        {\it University of Maryland, College Park, MD 20742}\\[.1in]
        {#3}\\[.1in] {\it {#4}}\\ {\it {#5}}\\[.4in]
        Physics Publication \#{#6}\\ {#7}\\[.5in] {\bf ABSTRACT}\\[.1in]
        \end{center} \begin{quotation}}                 % " for 2 authors
\def\border{                                            % border
        \setlength{\unitlength}{1mm}
        \newcount\xco
        \newcount\yco
        \xco=-24
        \yco=12
        \begin{picture}(140,0)
        \put(\xco,\yco){$\ktl$}
        \advance\yco by-1
        {\loop
        \put(\xco,\yco){$\kcl$}
        \advance\yco by-2
        \ifnum\yco>-240
        \repeat
        \put(\xco,\yco){$\kbl$}}
        \xco=158
        \yco=12
        \put(\xco,\yco){$\ktr$}
        \advance\yco by-1
        {\loop
        \put(\xco,\yco){$\kcr$}
        \advance\yco by-2
        \ifnum\yco>-240
        \repeat
        \put(\xco,\yco){$\kbr$}}
        \put(-20,11){\tiny University of Maryland Elementary Particle
Physics University of Maryland Elementary Particle Physics University of
Maryland Elementary Particle Physics}
        \put(-20,-241.5){\tiny University of Maryland Elementary
Particle Physics University of Maryland Elementary Particle Physics
University of Maryland Elementary Particle Physics}
        \end{picture}
        \par\vskip-8mm}
\def\bordero{                                           % alternate border
        \setlength{\unitlength}{1mm}
        \newcount\xco
        \newcount\yco
        \xco=-24
        \yco=12
        \begin{picture}(140,0)
        \put(\xco,\yco){$\ktl$}
        \advance\yco by-1
        {\loop
        \put(\xco,\yco){$\kcl$}
        \advance\yco by-2
        \ifnum\yco>-240
        \repeat
        \put(\xco,\yco){$\kbl$}}
        \xco=158
        \yco=12
        \put(\xco,\yco){$\ktr$}
        \advance\yco by-1
        {\loop
        \put(\xco,\yco){$\kcr$}
        \advance\yco by-2
        \ifnum\yco>-240
        \repeat
        \put(\xco,\yco){$\kbr$}}
        \put(-20,12){\ooo
bacdefghidfghghdhededbihdgdfdfhhdheidhdhebaaahjhhdahbahgdedgehgfdiehhgdigicba}
        \put(-20,-241.5){\ooo
ababaighefdbfghgeahgdfgafagihdidihiidhiagfedhadbfdecdcdf
agdcbhaddhbgfchbgfdacfediacbabab}
        \end{picture}
        \par\vskip-8mm}
\def\headpic{                                           % UM heading
        \indent
        \setlength{\unitlength}{.4mm}
        \thinlines
        \par
        \begin{picture}(29,16)
        \put(165,16){\line(1,0){4}}
        \put(170,16){\line(1,0){4}}
        \put(180,16){\line(1,0){4}}
        \put(175,0){\line(1,0){4}}
        \put(180,0){\line(1,0){4}}
        \put(185,0){\line(1,0){4}}
        \put(169,0){\line(0,1){16}}
        \put(170,0){\line(0,1){16}}
        \put(179,0){\line(0,1){16}}
        \put(180,0){\line(0,1){16}}
        \put(184,0){\line(0,1){16}}
        \put(185,0){\line(0,1){16}}
        \put(169,16){\oval(8,32)[bl]}
        \put(170,16){\oval(8,32)[br]}
        \put(179,0){\oval(8,32)[tl]}
        \put(185,0){\oval(8,32)[tr]}
        \end{picture}
        \par\vskip-6.5mm
        \thicklines}
\def\title#1#2#3#4{\border\headpic {\hbox to\hsize{#4 \hfill UMDEPP #3}}\par
        \begin{center} \vglue .5in {\large\bf #1}\\[.6in]
        {#2}\\[.1in] {\it Department of Physics and Astronomy}\\
        {\it University of Maryland, College Park, MD 20742}\\[1.5in]
        {\bf ABSTRACT}\\[.1in] \end{center} \begin{quotation}}  % title stuff
\def\Title#1#2#3#4#5#6#7{\border\headpic
        {\hbox to\hsize{#7 \hfill UMDEPP #6}}\par
        \begin{center} \vglue .4in {\large\bf #1}\\[.4in]
        {#2}\\[.1in] {\it Department of Physics and Astronomy}\\
        {\it University of Maryland, College Park, MD 20742}\\[.1in]
        {#3}\\[.1in] {\it {#4}}\\ {\it {#5}}\\[.5in] {\bf ABSTRACT}\\[.1in]
        \end{center} \begin{quotation}}                 % " for 2 authors
\def\endtitle{\end{quotation}\newpage}                  % end title page

\def\sect#1{\bigskip\medskip \goodbreak \noindent{\bf {#1}} \nobreak \medskip}
\def\refs{\sect{REFERENCES} \footnotesize \frenchspacing \parskip=0pt}
\def\Item{\par\hang\textindent}

\begin{document}
% ================ End of def.tex ======================
%``

\def\gg{{\hbox{\sc g}}}
\def\nt{$~N=2$~}
\def\gg{{\hbox{\sc g}}}
\def\nt{$~N=2$~}
\def\tr{{\rm tr}}
\def\Tr{{\rm Tr}}
\def\mpl#1#2#3{Mod.~Phys.~Lett.~{\bf A{#1}} (19{#2}) #3}

\def\scst{\scriptstyle}
\def\itrema{$\ddot{\scriptstyle 1}$}

\def\ula{{\underline a}} \def\ulb{{\underline b}} \def\ulc{{\underline c}}
\def\uld{{\underline d}} \def\ule{{\underline e}} \def\ulf{{\underline f}}
\def\ulg{{\underline g}}

\def\plpl{{+\!\!\!\!\!{\hskip 0.009in}{\raise -1.0pt\hbox{$_+$}}
{\hskip 0.0008in}}}

\def\mimi{{-\!\!\!\!\!{\hskip 0.009in}{\raise -1.0pt\hbox{$_-$}}
{\hskip 0.0008in}}}

\def\items#1{\\ \item{[#1]}}
\def\ul{\underline}
\def\un{\underline}
\def\-{{\hskip 1.5pt}\hbox{-}}

\def\fracmm#1#2{{{#1}\over{#2}}}

\def\footnotew#1{\footnote{\hsize=6.5in {#1}}}

\def\low#1{{\raise -3pt\hbox{${\hskip 0.75pt}\!_{#1}$}}}

\def\ip{{=\!\!\! \mid}}
\def\ze{\zeta^{+}}
\def\zeb{{\bar \zeta}^{+}}
\def\umb{{\underline {\bar m}}}
\def\unb{{\underline {\bar n}}}
\def\upb{{\underline {\bar p}}}
\def\um{{\underline m}}
\def\up{{\underline p}}
\def\Phib{{\Bar \Phi}}
\def\Phit{{\tilde \Phi}}
\def\Phibt{{\tilde {\Bar \Phi}}}
\def\Db{{\Bar D}_{+}}
\def\gg{{\hbox{\sc g}}}
\def\nt{$~N=2$~}

\border\headpic {\hbox to\hsize{February 1992 \hfill UMDEPP 92-171}}\par
\begin{center}
\vglue .25in

{\large\bf Supersymmetric ~Self-Dual~ Yang-Mills~ and
{}~Supergravity \\
as ~Background ~of ~Green-Schwarz ~Superstring$\,$\footnote{Research supported
by NSF grant \# PHY-91-19746.} }\\[.1in]

\baselineskip 10pt

\vskip 0.25in

Hitoshi NISHINO ~ and ~S.~James GATES, Jr. \\[.1in]
{\it Department of Physics} \\ [.015in]
{\it University of Maryland at College Park}\\ [.015in]
{\it College Park, MD 20742-4111, USA} \\[.1in]
and\\[.1in]
{\it Department of Physics and Astronomy} \\[.015in]
{\it Howard University} \\[.015in]
{\it Washington, D.C. 20059, USA} \\[.18in]
and
\\[.18in]
Sergei V.~KETOV$\,$\footnote{On leave of absence from High Current Electronics
Institute of the Russian Academy of Sciences, Siberian Branch,
Akademichesky~4, Tomsk 634055, Russia.} \\[.1in]
{\it Department of Physics} \\[.015in]
{\it University of Maryland at College Park}\\[.015in]
{\it College Park, MD 20742-4111, USA} \\[.001in]

\vskip 0.35in

{\bf ABSTRACT}\\[.1in]
\end{center}

\begin{quotation}

{\baselineskip 5pt

We present the Green-Schwarz $~\s\-$model coupled to
the $~N=1$~ {\it {supersymmetric}}
Yang-Mills and supergravity in a
four-dimensional space-time with the indefinite signature $(+,+,-,-)$.
We first confirm the $~\k\-$invariance of the
Green-Schwarz action, and show that all the $~\b\-$functions for the
backgrounds vanish consistently after the use of their superfield equations.
Subsequently, we inspect the supersymmetric {\it self-duality}
conditions, that have been developed in our previous paper on the
Yang-Mills and supergravity backgrounds.  Remarkably, the Majorana-Weyl
spinor dictating the {\it supersymmetric self-duality} conditions is
consistent with the couplings of Green-Schwarz superstring.
Such Green-Schwarz superstring is supposed to be the underlying theory
of the {\it supersymmetric self-dual} Yang-Mills theory, which is conjectured
to generate {\it all} exactly soluble supersymmetric systems in
lower dimensions.
}

\endtitle
% ============== csdef.tex =====================
\def\doit#1#2{\ifcase#1\or#2\fi}
\def\[{\lfloor{\hskip 0.35pt}\!\!\!\lceil}
\def\]{\rfloor{\hskip 0.35pt}\!\!\!\rceil}
\def\delsl{{{\partial\!\!\! /}}}
\def\caldsl{{\calD\!\!\! /}}
\def\calO{{\cal O}}
\def\asym{({\scriptstyle 1\leftrightarrow \scriptstyle 2})}
\def\Lag{{\cal L}}
\def\du#1#2{_{#1}{}^{#2}}
\def\ud#1#2{^{#1}{}_{#2}}
\def\dud#1#2#3{_{#1}{}^{#2}{}_{#3}}
\def\udu#1#2#3{^{#1}{}_{#2}{}^{#3}}
\def\calD{{\cal D}}
\def\calM{{\cal M}}
\def\tildef{{\tilde f}}
\def\calDsl{{\calD\!\!\!\! /}}

\def\Hat#1{{#1}{\large\raise-0.02pt\hbox{$\!\hskip0.038in\!\!\!\hat{~}$}}}
\def\hati{{\hat{I}}}
\def\dt{$~D=10$~}
\def\alp{\alpha{\hskip 0.007in}'}
\def\oalp#1{\alp^{\hskip 0.007in {#1}}}
\def\naive{{{na${\scriptstyle 1}\!{\dot{}}\!{\dot{}}\,\,$ve}}}
\def\items#1{\vskip 0.05in\Item{[{#1}]}}
\def\item#1{\Item{#1}}

\def\pl#1#2#3{Phys.~Lett.~{\bf {#1}B} (19{#2}) #3}
\def\np#1#2#3{Nucl.~Phys.~{\bf B{#1}} (19{#2}) #3}
\def\prl#1#2#3{Phys.~Rev.~Lett.~{\bf #1} (19{#2}) #3}
\def\pr#1#2#3{Phys.~Rev.~{\bf D{#1}} (19{#2}) #3}
\def\cqg#1#2#3{Class.~and Quant.~Gr.~{\bf {#1}} (19{#2}) #3}
\def\cmp#1#2#3{Comm.~Math.~Phys.~{\bf {#1}} (19{#2}) #3}
\def\jmp#1#2#3{Jour.~Math.~Phys.~{\bf {#1}} (19{#2}) #3}
\def\ap#1#2#3{Ann.~of Phys.~{\bf {#1}} (19{#2}) #3}
\def\prep#1#2#3{Phys.~Rep.~{\bf {#1}C} (19{#2}) #3}
\def\ptp#1#2#3{Prog.~Theor.~Phys.~{\bf {#1}} (19{#2}) #3}
\def\ijmp#1#2#3{Int.~Jour.~Mod.~Phys.~{\bf {#1}} (19{#2}) #3}
\def\nc#1#2#3{Nuovo Cim.~{\bf {#1}} (19{#2}) #3}
\def\ibid#1#2#3{{\it ibid.}~{\bf {#1}} (19{#2}) #3}

\def\szet{{${\scriptstyle \b}$}}
\def\ula{{\un a}}
\def\ulb{{\un b}}
\def\ulc{{\un c}}
\def\uld{{\un d}}
\def\ulA{{\un A}}
\def\ulM{{\underline M}}
\def\cdm{{\Sc D}_{--}}
\def\cdp{{\Sc D}_{++}}
\def\vTheta{\check\Theta}
\def\Pisl{{\Pi\!\!\!\! /}}

\def\fracmm#1#2{{{#1}\over{#2}}}
\def\gg{{\hbox{\sc g}}}
\def\half{{\fracm12}}
\def\ha{\half}

\def\fracm#1#2{\hbox{\large{${\frac{{#1}}{{#2}}}$}}}
\def\Dot#1{{\hskip 0.3pt}{\raise 7pt\hbox{\large .}}
\!\!\!\!{\hskip 0.3pt}{#1}}
% ============== End of csdef.tex ========================
\def\Dot#1{\buildrel{_{_\bullet}}\over{#1}}

\oddsidemargin=0.03in
\evensidemargin=0.01in
\hsize=6.5in
\textwidth=6.5in

\noindent 1. {\it Introduction.}
{}~~~It is a commonly accepted conjecture that self-dual Yang-Mills
(SDYM) theories in four-dimensions with the indefinite metric $~(+,+,-,-)$~
[1]\footnotew{We call this $~D=(2,2)$~ space-time.  In the context with
{\it no} need of distinction of signature, the expression $~D=4~$ is
also used.} generate {\it all}
possible exactly soluble (bosonic) models in lower dimensions after some
suitable compactifications or dimensional reductions [2].
This rather ``mathematical'' conjecture has been recently extended to
make a connection with string theory by the observation [3] that
self-duality (SD) is required for the Yang-Mills (YM) fields as the consistent
massless physical vectorial fields in the $~N=2$~ superstring, based on
amplitude calculations.  Moreover the fact that the critical dimension
of the $~N=2$~ superstring is exactly $~D=(2,2)$~ [4] (in real coordinates)
also indicates that the SDYM theory is the consistent backgrounds for the
heterotic $~N=(2,0)$, $~N=(2,1)$ and the open $~N=(2,2)$~ superstrings.  In
other words, the SDYM theory in $~D=(2,2)$, generating {\it
all} exactly soluble (bosonic) models, is itself the {\it low energy
effective theory} of these various different superstrings.

In our previous paper [5] we have shown in a covariant way that the
$~N=(2,0)~$ Neveu-Schwarz-Ramond (NSR) superstring [6] can consistently couple
{\it only} to SDYM background in $~D=(2,2)$,
using the NSR $~\s\-$model [7], in the absence of gravity background.
We have concluded that the
condition of vanishing $~\b\-$functions for general YM
backgrounds restricts the YM backgrounds to be {\it self-dual}.

In the NSR formulation, the background fields did {\it not} possess
manifest {\it supersymmetry}.  This is due to the usual fate of the NSR
superstring, in which the supersymmetry is manifest {\it only} on the
world-sheet [6].  Motivated by this result, it is natural to consider the
possibility of {\it supersymmetric} YM backgrounds coupled to Green-Schwarz
(GS) $~\s\-$model, in which the target space-time has {\it manifest}
supersymmetry.  In contrast to the {\it non-supersymmetric} SDYM theory, the
{\it supersymmetric} SDYM system can be assumed to create all the exactly
soluble {\it supersymmetric} systems in lower-dimensions.

In our recent paper [8], we have shown how to construct the {\it supersymmetric
self-dual} YM and supergravity (SG) multiplets in the space-time with
the signature $~(+,+,-,-)$.\footnotew{We stress that the signature
$~(+,+,-,-)$~ of our space-time is crucial
for generating soluble systems in lower dimensions.  Hence the previous
trials [9] based on $~D=(1,3)~$ are {\it not} appropriate for our
purpose.}~~  We have found that there exist
Majorana-Weyl spinors in such space-time, that dictate the {\it
supersymmetric} SD condition for the YM and SG multiplets.

In this paper we apply the results of our recent paper [8] on {\it
supersymmetric} SD conditions to the constructions of GS
superstring, that can consistently couple to the background SDYM and
SDSG\footnotew{As we will see later, the phrase ``SG multiplet''
also includes a tensor multiplet $~(B_{m n},\,\chi,\,\Phi)$.}~~We first
introduce the $~N=1$~ YM and SG backgrounds in {\it curved} superspace,
yet {\it without} imposing the SD conditions on the backgrounds.
We next formulate the action principle for the
$~N=(2,0)$ heterotic GS superstring, coupled to $~D=(2,2),~N=1$~
supersymmetric YM plus SG background.
Accordingly, we inspect the $~\k\-$invariance of the GS
action, and compute the $~\b\-$functions for the YM plus SG backgrounds.
We finally examine the most important aspect, namely the
consistency of SD conditions on the YM and SG backgrounds with the vanishing
$~\b\-$functions.

        The notations we are using is the same as our previous paper
[8], in order to save space for notational explanation.  We heavily rely
also on the important conceptual results of the previous paper [8].

\bigskip\bigskip

\noindent 2. {\it Superspace for YM plus SG.}~~~We first
introduce the YM and SG multiplets in {\it curved} superspace.
In this section, we do {\it not} impose any SD
condition, maintaining the YM and SG multiplet more general.  This is
always possible, because the SD conditions are understood as creating
the ``sub-multiplet'' of the usual YM and SG multiplet, as we explained
in our paper [8].

As the most important features in our notation [8] for the spinorial
manipulations, we mention the following property:  For any inner product
in terms of two Majorana spinors $~\psi$~ and $~\chi$, we can simply
perform the following replacement in order to switch on to the
two-component Weyl representation.  For example,
$$\eqalign{\Bar\psi\G^\ula\chi~ \rightarrow ~ & +\psi \s^\ula \Bar\chi
+ \Bar\psi \Tilde\s ^\ula \chi \cr
\equiv\, & + \psi^\a (\s^\ula)\du \a {\Dot \b} \Bar\chi_{\Dot\b}
+\Bar\psi^{\Dot\a} ({\Tilde\s}^\ula)\du {\Dot\a} \b \chi{\low\b}
= + \psi^\a (\s^\ula)\du\a{\Dot\b} \Bar\chi_{\Dot\b} + \Bar\psi^{\Dot\a}
(\s^\ula)\ud\b{\Dot\a} \chi{\low \b} ~~. \cr }
\eqno(2.1) $$
Here the most left side is in terms of {\it four-component} Majorana
spinors, while other sides are in terms of {\it two-component} Weyl
spinor notations [8].  Due to the parallel feature between the {\it dotted}
and {\it undotted} indices, we sometimes omit the ``dot-conjugate''
expressions to save space.

Since the total system is already complicated, we give the
coupled system at one stroke here, instead of separating them.
The superfield equations for the SG multiplet $~(e\du {\underline m} \ula,
\psi\du{\underline m}{\underline \m})~\oplus$
$(B_{\underline m\underline n},
\chi_{\underline \a},\Phi)~$ coupled to the YM multiplet ~$(A\du
{\underline m} I,\l\du{\underline\a} I)$~ are obtained as usual from
Bianchi identities (BIds.) [10], which
are\footnotew{We give here a brief note for our notations [8]
as a quick reference.  For the vectorial indices for the local Lorentz frame
we use the {\it underlined} indices $~{\scst
\ula\,\equiv\,(a,~\Bar a),~\ulb\,\equiv\,(b,~\Bar b),~ \cdots}$, which
are equivalent to the pairs of 2-dimensional complex coordinate indices
$~{\scst a,~b,~\cdots= 1,~2}$~ and their complex
conjugates $~{\scst\Bar a,~\Bar b,~\cdots\,=\,\Bar1,~\Bar 2}$.
As for the spinorial indices, we use the {\it undotted} and {\it dotted}
indices $~{\scst\a,~\b,~\cdots\,=\, 1,~2~}$ and $~{\scst\Dot
\a,~\Dot\b,~\cdots\,=\,\Dot 1,~\Dot 2}$~ for Weyl spinors of each
chirality.  We also use the {\it underlined} spinorial indices $~{\scst
\ul\a\,\equiv\,(\a,~\Dot\a),~\ul\b\,\equiv\,(\b,~\Dot\b),~ \cdots\,}$,
denoting the pairs of {\it dotted} and {\it undotted} indices.}
$$\eqalign{&\nabla_{\[ A} T\du{B C)} D - T\du{\[ A B|} E T\du {E|C)} D -
\half R\du{\[A B C)\,d} e {\cal M}\du ed \equiv 0~~, \cr
&\fracm 16 \nabla_{\[ A} G_{B C D)} - \fracm 14 T\du {\[ A B|} E G_{E|C
D)}
\equiv \fracm 14 F\du{\[ A B} I F\du {C D)}  I ~~, \cr
&\half \nabla_{\[ A} F\du{B C)} I - \fracm 1 4 T\du{\[ A B|} D F\du{|D C)} I
\equiv 0 ~~. \cr }
\eqno(2.2) $$
The indices $~{\scriptstyle I,~J,~\cdots}$~ are for the adjoint indices
of the YM group.
The r.h.s.~of the $~\nabla G\-$BIds.~need the
YM Chern-Simons term, as in the usual SG theories
[11].\footnotew{The ~$G_{\ul m \ul n \ul p}$~ is the field strength of
$~B_{\ul m \ul n}$.
There may also arise at higher-order in $~\a\,'$~ the Lorentz
Chern-Simons form, which we do not treat in this paper.}
We next have to solve these BIds.~by imposing some
lower-dimensional constraints at $~0\le d\le 1 $~
(in mass dimensions).  In order also to
simplify the computation of $~\b\-$functions later, we use what is
called ``$\b\-$function favoured constraints'' (BFFC) originally
developed in Ref.~[12] for $~D=10$~ GS $~\b\-$function calculation [13] :
$$\eqalign{&T\du{\a\Dot\b} \ulc = i(\s^\ulc)_{\a\Dot\b} ~~, ~~~~
\nabla_\a \Phi = -\fracm 1{\sqrt 2} \chi_\a ~~, \cr
&G_{\a\Dot\b \ulc} = i\half (\s_\ulc) _{\a\Dot\b} ~~,~~~~ G_{\a\b\ulc} =
0~~, ~~~~ G_{\ul\a\ulb\ulc} = 0~~, ~~~~ G_{\ul\a\ul\b\ul\g}=0~~, \cr
&T\du{\a\b} {\Dot\g} = 0~~, ~~~~T\du{\a\Dot\b} \g = {\sqrt 2} \d\du\a\g
\Bar\chi_{\Dot \b} ~~, ~~~~ T\du{\a\b} \g = -{\sqrt 2}\d\du {(\a}\g
\chi{\low{\b)}} ~~, \cr
& \nabla_\a\Bar\chi_{\Dot\b} = -i\fracm 1{\sqrt 2} (\s^\ulc) _{\a\Dot\b}
\nabla_\ulc \Phi + i \fracm 1{12{\sqrt2}}(\s^{\ulc\uld\ule})_{\a\Dot\b}
G_{\ulc\uld\ule} - \fracm 1 {4{\sqrt 2}} (\s^\ulc)_{\a\Dot\b} (2\chi
\s_\ulc\Bar\chi-\l^J\s_\ulc \Bar\l^J) ~~, \cr
& T\du{\a\ulb}\g =-i \fracm 1 8 (\s_\ulb\s^\ulc)\du\a\g
(\l^I\s_\ulc \Bar\l^I)  ~~, ~~~~T\du{\ul\a\ulb}\ulc = 0~~, \cr
& R_{\a\Dot\b c d} = -2i (\s^\ule)_{\a\Dot\b} G_{\ulc\uld\ule} ~~, \cr
& F\du{\a\ulb} I =i \fracm 1{\sqrt 2} (\s_\ulb \Bar\l^I) _\a ~~, \cr
&\nabla_\a\l^{\b\, I} =- \fracm 1{2\sqrt2} (\s^{\ulc\uld})\du\a\b
F\du{\ulc\uld} I  - \fracm 1{2{\sqrt 2}} (\s^{\ulc\uld})\du \a\b
(\Bar\chi\s_{\ulc\uld} \l^I)
-{\sqrt 2} \chi_\a\l^{\b\, I} -\fracm 1{\sqrt 2} \d\du\a\b
(\chi\l{}^I+\Bar\chi \Bar\l^I)  ~~, \cr
& \nabla_\a\Bar\l^{\Dot\b\, I} = -{\sqrt 2} \chi_\a\Bar\l ^{\Dot\b\, I}
{}~~. \cr}
\eqno(2.3) $$
The advantage of this set of constraints is the simplification for the
one-loop $~\b\-$function [12,13].

Out of the higher-dimensional ~$(d\ge 3/2)$~ BIds.~we
get the superfield equations for the
\newpage
\noindent SG multiplet coupled to the
YM multiplet:
$$\li{&R_{(\ula\ulb)} + 4 \nabla_{(\ula} \nabla_{\ulb)} \Phi
- 4F\du \ula {\ulc I} F\du{\ulb\ulc} I
+ i\l^I \s_{(\ula}\nabla_{\ulb)} \Bar\l{} ^I + i\Bar\l{}^I
\s_{(\ula} \nabla_{\ulb)} \l^I = 0 ~~,
&(2.4) \cr &~~~  \cr
& \nabla_{\ulc} G\du{\ula\ulb} \ulc
+ 4 G\du{\ula\ulb} \ulc \nabla_{\ulc} \Phi - {\sqrt 2} \chi T_{\ula\ulb}
-{\sqrt 2}\Bar\chi\Bar T_{\ula\ulb}
+ i\half \Bar\l{}^I \s_{\[ \ula} \nabla_{\ulb\]} \l^I + i\half
\l^I \s_{\[\ula} \nabla_{\ulb \]} \Bar\l{}^I  \cr
& ~~~~~ + 4(\chi\l^I +\Bar\chi\Bar\l{}^I ) F\du{\ula\ulb} I = 0 ~~,
&(2.5) \cr
& ~~~~ \cr
& i(\s^\ulb \Bar T_{\ula\ulb} ) _\a - 2{\sqrt 2} \nabla_\ula\chi_\a
-{\sqrt 2}i (\s^\ulb\Bar \l^I) _\a F\du{\ula\ulb} I = 0 ~~,
&(2.6) \cr
& ~~~~ \cr
& i(\s^\ula \nabla_\ula \Bar\l{}^I)_\a + 4i (\s^\ula\Bar\l{}^I)_\a
\nabla_\ula \Phi + 2{\sqrt2} \nabla_\a (\chi\l^I +\Bar\chi\Bar\l{}^I)
+\half (\s^\ulc\Bar\l{}^I) _\a(\l^J\s_\ulc \Bar\l^J) = 0 ~~,
&(2.7) \cr
& ~~~~ \cr
& \nabla_\ulb F\du\ula{\ulb \,I} + 4F\du\ula{\ulb\, I} \nabla_\ulb \Phi \cr
& ~~~~~ + \bigg[- \fracm i2 f^{I J K} (\l^J \s_\ula \Bar\l^K)
+2\nabla_\ula(\chi\l^I) +i \fracm 1{\sqrt2} (T_{\ula\ulb} \s^\ulb \Bar\l^I )
\cr
& ~~~~~ ~~~~~ ~~ + i \fracm 32 (\l^I \s^\ulb \Bar\l^J ) F\du{\ula\ulb}J
-i \fracm 14 (\l^I \s^{\ula\ulb\ulc} \Bar\l^J) F\du{\ulb\ulc} J
&(2.8) \cr
& ~~~~~ ~~~~~ ~~ + \fracm 14 \chi\du\ula{\ulb\ulc} F\du{\ulb\ulc} I
+ i \fracm 3{16} (\chi \s_\ula \s^\ulb \l^I) (\l^J \s_\ulb \Bar\l^J)
+\fracm1{16} (\chi \s^{\ulb\ulc} \l^I ) \l_{\ula\ulb\ulc}
+ ({\scst\rm dotted}~\leftrightarrow~{\scst\rm undotted}) \bigg] = 0 ~~,
\cr } $$
where $~\l_{\ula\ulb\ulc} \equiv i(\l^I\s_{\ula\ulb\ulc} \Bar\l^I)$
$\equiv i \l^{\a\,I} (\s_{\ula\ulb\ulc})\du{\a}{\Dot\b}
\Bar\l_{\Dot\b}{}^I$.
As is easily noticed, due to the similarity of our $~D=(2,2)$~ system to
the usual $~D=(1,3)$~ case, all of  these computations are made simpler
in terms of {\it underlined} vectorial indices.

\bigskip\bigskip

\noindent 3. {\it GS ~$\s\-$Model and $~\k\-$Symmetry.}~~~
We can also formulate the GS $~\s\-$model, once we have understood the
superspace structure on the hermitian manifold in $~D=(2,2)$.
Its form is formally the same as the usual GS $~\s\-$model
in $~D=(1,3)$ ~[14], as long as we are using the {\it underlined}
indices [8].  Our GS action is ~$I = I_{\rm SG} + I_{\rm YM}$, where
$$\li{& I_{\rm SG} \equiv \int d^2 \s ~ \left[\,\half e g^{i j}
\eta_{\ula\ulb}\Pi\du i\ula \Pi\du j \ulb + \e^{ij}\, \Pi\du i B \Pi\du j
A B_{A B} \right] \equiv \int d^2\s ~\Lag_{\rm SG} ~~,
& (3.1)  \cr
&I_{\rm YM} \equiv \int d^2 \s ~\left[ \, \half i e \Bar\psi_-\g^i \left(
\partial_i\psi_- + \Pi\du i A A\du A I T^I \psi_- \right) \,\right] \equiv \int
d^2\s\, \Lag_{\rm YM} ~~.
& (3.2) \cr} $$
As usual $~\Pi\du iA\equiv(\partial_i Z^M) E\du MA~$ is with the
supervielbein in the target $~D=(2,2),~N=1$~ superspace with the
supercoordinates $~Z^M$.  The $~e$~ is the determinant of the
two-dimensional $~(d=2)$~
zweibein, the spinors $~\psi_-$~ are $~d=2$~unidextrous
Majorana-Weyl spinors, and $~\g{\low i}$~ is the gamma-matrices on
the world-sheet, and
$~{\scriptstyle i,~j,~\cdots}$~ are for its curved coordinates.
The background superfield $~B_{M N}$~ is a part of the tensor multiplet, which
is conveniently regarded as a part of the ``SG multiplet'' in this paper.
The $~A\du MI$~ is the background YM superfield, coupled by the
unidextrous fermions $~\psi_-$.

        As in the usual GS $\s\-$model [6], this action has the
$~\k\-$symmetry dictated by
$$\eqalign{ &\d_\k E^{\ul\a} \equiv (\d_\k Z^M) E\du M{\ul\a}
= i \Pi\du \plpl\ula (\s_\ula)^{\ul\a\ul\b} \k_{\mimi \ul\b}~~, \cr
& \d_\k E^\ula \equiv (\d_\k Z^M) E\du M\ula = 0 ~~, \cr
&\d_\k e\du \mimi i = 4i \left( \k_\mimi{}^{\ul\a} \Pi_{\mimi\ul\a}
\right) e\du\plpl i
+ i \fracm 1{2{\sqrt 2}} (\k\du \mimi {\ul\a} \l\du{\ul\a} I) (\Bar\psi_- T^I
\psi_- ) e\du \plpl i ~~, \cr
&\d_\k e\du \plpl i = 0 ~~, \cr
&\d_\k \psi \du - r = -(\d_\k E^{\ul \g}) A\du {\ul\g} I (T^I\psi_-)^r
{}~~, \cr}
\eqno(3.3) $$
where we use the world-sheet light-cone projections $~{\scriptstyle
\plpl}~$ and ~${\scriptstyle\mimi}$~
defined by $~(\eta^{i j} \pm \e^{ij})/2$, respectively.  The indices
$~{\scriptstyle r,~s,~ \cdots} ~$ are used for the representation of the
unidextrous fermions $~\psi_-^r$.  We stress here that
owing to the BFFC [12,13], these transformation rules are much
simplified
compared with other constraints.  Since the basic
structure is parallel to the usual GS case, we skip other explanations.

\bigskip\bigskip

\noindent 4. {\it $~\b\-$Function Calculations.}~~~We next
see how the condition of vanishing of $~\b\-$functions are consistent
with the superfield equations of YM and SG multiplets.  To this end we
derive the $~\b\-$functions for the YM multiplet up to the three-loop,
while for the SG multiplet up to the two-loop level in the $~\s\-$model
perturbation.
Eventually we will get the quadratic order terms in the YM multiplet in
the SG superfield equations.\footnotew{As in
the $~D=10$~ case [13], we skip the two-loop contribution from the
pure SG.}~~As a guiding reference to save space, we
refer the readers to Ref.~[13], where the GS $~\s\-$model in $~D=10$~ was
treated.

        As in the $~D=10$~ case [13], we need the classical-quantum
splitting to perform the $\b\-$function calculation.  To this end, we
reply on the technique of $~\D\-$operation which extract the quantum
fluctuation from the original GS lagrangians (3.1) and (3.2).
For example, we have
$$\eqalign{ &\D \Pi\du i A \equiv D_i\xi ^A- \Pi\du i D \xi^B
T\du{B D} A~~, ~~~~
D_i\xi^A \equiv \partial_i \xi^A -\Pi\du iD \phi\du D {A B}
\xi_B~~, \cr
&\D(D_i\xi^E) \equiv \xi^D \Pi\du i B R\du{ABD} E~~, ~~~~
\D B_{A B} = \xi^C \nabla_C B_{A B} ~~,  \cr}
\eqno(4.1) $$
where $~\xi^A $~ are the quantum fluctuation of the superspace
coordinates $~Z^M$.  After successive $~\D\-$operations. we can get any
terms at arbitrary $~n\-$th order lagrangian $~\Lag^{(n)}$~
in $~\xi^A$.  There is also some treatment related to the
gauge non-covariant terms created
by the $~\D\-$operation out of the YM unidextrous lagrangian
$~\Lag_{\rm YM}$, which is understood by considering the measure-change in
the fermionic path-integral under the gauge transformation.  We skip all
the details here, but for the details see Refs.~[13,14].  After all, we get
the relevant lagrangian quadratic in $~\xi^A$:
%\newpage
$$\li{ e^{-1}\Lag^{(2)}_{\rm SG} = \,&\half (D^i\xi^\ula) (D_i\xi_\ula)
+ \Pi\du + \ula (D_-\xi)^{\ul\a} (\s_\ula \xi) _{\ul\a}
-2(D_\mimi \xi^\ula) \xi^\ulb \Pi \du \plpl \ulc G_{\ula\ulb\ulc} \cr
& -\half \xi^\ula \xi^\ulb \Pi\du i\ulc \Pi^{i\,\uld} R_{\ulb\ulc\uld
\ula} + \half \e^{ij} \Pi\du i\ulc \Pi\du j\ulb
\xi^\ula\xi^\uld\nabla_\uld G_{\ula\ulb\ulc}
&(4.2)  \cr
&- i\half \xi^\ula \xi^\uld\xi^\ulb \Pi\du i \ulc \Pi^{i\,\ul\d}
(\s_{\[\ula}  T_{\ulc \] \ulb} )_{\ul\d} - i \half \e^{ij} \xi^\ule \xi
^\ulc \Pi\du i\uld \Pi\du j{\ul\a} (\s_\ule T _{\ulc\uld} )
_{\ul\a} + (\hbox{terms with}~ \xi^{\ul\a})~~,\cr} $$
where we omitted the $~(\xi^A)^2\-$terms with $~\xi^{\ul\a}$~ except for
the ``kinetic term'' of $~\xi^{\ul\a}$~ (i.e., the second term),
because they
do not contribute to our $\b\-$fucntions, for the reason to be mentioned
shortly.  Similarly the unidextrous lagrangian is
$$\li{e^{-1}\Lag_{\rm YM}^{(2)} =& \,\fracm i2 {\Bar\Psi}_-
\partial_\plpl \Psi_-
+ \fracm i2 {\Bar\Psi}_- \Pi\du \plpl A A_A{}^I T^I \Psi_-
+i\Bar\psi_-\Pi\du \plpl A \xi^B F_{B A}{}^I T^I \Psi_- \cr
&\, +\fracm i4 {\Bar\psi}_- \left[ \, (D_\plpl\xi^A + \Pi \du \plpl
B\xi^C T\du{C B}A ) \xi ^D
F_{DA}{}^I + \Pi\du \plpl A\xi^B \xi^C (\nabla_C F_{B A}{}^I )
\,\right] T^I \psi_- ~~.
&(4.3) \cr} $$
Here we are splitting also the fermion $~\psi_-\rightarrow \psi_- +
\Psi_-$~ for the classical value $~\psi_-$~ and the quantum value
$~\Psi_-$.

        The $\b\-$function calculation is much parallel to the $~D=10$~
GS case [13], as well as to the $~D=4$~ GS case [14], so that we do not
repeat the details, except the main results.  The relevant graphs to be
computed for the one-loop $\b\-$functions for the SG is a tadpole
graph like Fig.~1(a) in [13].  This is why we could ignore the
$~\xi^{\ul\a}\-$terms in (4.2), since these terms do
{\it not} generate any $~1/\e\-$pole in such a tadpole graph [13].  This
is true also for other graphs in this paper.   For the YM
multiplet, we have three
graphs to be evaluated like Fig.~2 in Ref.~[13].  Owing to the BFFC,
these graph calculations are drastically simplified [13].
At the two-loop level, we
evaluate only the YM corrections, namely the (YM)$^2\-$type corrections
in the SG superfield equations.  The relevant graphs are the $~D=4$~
counterpart of the four graphs in Fig.~3 in Ref.~[13].

        As usual, the $\b\-$functions are obtained as the $~1/\e\-$pole
parts of these graphs.  We also need to perform some field redefinitions,
namely the change of string valuable $~\d Z^M = \e^M(Z)$, and the
fermion chiral rotation:
$~\psi \rightarrow \exp(\L^I T^I) \psi$,
$$\e^\ula =\fracmm 2 \e \nabla^\ula \Phi ~~, ~~~~
\e^{\ul\a} = \fracmm 1 {2\e} \chi^{\ul\a} (\l^I \chi
+ \Bar\l^I \Bar\chi) ~~,
{}~~~~ \L^I = \fracmm1 \e (\l^I\chi+ \Bar\l^I\Bar\chi)
{}~~.
\eqno(4.4) $$
After these field redefinitions we get the $~1/\e\-$pole parts for the
one-loop YM $~\b\-$function:
$$\li{e^{-1} \Lag_{\rm YM}^{\infty(L=1)} = \fracmm1 \e i (\Bar\psi_-\g^i
T^I \psi_-) \bigg[ & \Pi\du i\ula\left\{ \nabla_\ulc F\udu\ulc\ula I
-(T_{\ulc\ula} \s^\ulc \Bar\l^I) - (\Bar T _{\ulc\ula}\s^\ulc \l^I)
- f^{I J K} (\l^J \s_\ula \Bar\l^K) \right\} \cr
& -\Pi\du i{\ul\a} (\s^\ulc \nabla_\ulc \l^I)
_{\ul\a} \bigg] ~~,
&(4.5) \cr} $$
and for the $~\b\-$functions for the SG multiplet, which has the
one-loop SG plus two-loop YM contributions:
\newpage
$$\li {&e^{-1} \Lag_{\rm SG}^{\infty(L=1)} = \fracmm1\e \bigg\{
\fracm14 \Pi\du i\ula \Pi^{i\ulb} \Big[ R_{(\ula\ulb)} + 4
\nabla_{(\ula}\nabla_{\ulb)} \Big] \cr
& ~~~~~ ~~~~~ ~~~~~ ~~~~~  - \half \e^{ij} \Pi\du i\ula \Pi\du j\ulb
\Big[\nabla_\ulc
G\du{\ula\ulb} \ulc + 4 G\du{\ula\ulb}\ulc \nabla_\ulc\Phi
-{\sqrt2} (\chi T_{\ula\ulb} + \Bar\chi\Bar T_{\ula\ulb} ) \Big] \cr
& ~~~~~ ~~~~~ ~~~~~ ~~~~~ - \Pi\du \plpl {\ul\b} \Pi\du \mimi{\ula}
\Big[ i\s_\ulb T\du\ula\ulb - 2{\sqrt2} \nabla_\ula\chi
\Big]_{\ul\b} \bigg\}~~,
&(4.6) \cr } $$
$$\li{&~~~~ \cr
&e^{-1}\Lag_{\rm SG}^{\infty(L=2)} = \fracmm1\e \bigg\{\Pi\du
i\ula \Pi^{i\ulb} \bigg[ 2  F_{\ula\ulc}{}^I F\udu\ulc\ulb I +
i \half \l^I \s_{(\ula} \nabla_{\ulb)} \Bar\l^I +
i \half  \Bar\l^I \s_{(\ula} \nabla_{\ulb)} \l^I \Big] \cr
& ~~~~~ ~~~~~ ~~~~~ ~~~~~ -\half \e^{ij} \Pi\du i\ula \Pi\du j\ulb \Big[
i \fracm14 \l^I \s_{\[\ula} \nabla_{\ulb\]} \Bar\l^I
+ i \fracm 14 \Bar\l^I \s_{\[\ula} \nabla_{\ulb\]} \l^I
+2 F\du{\ula\ulb} I (\l^I \chi+ \Bar\l^I \Bar\chi ) \Big] \cr
& ~~~~~ ~~~~~ ~~~~~ ~~~~~ + \e^{ij} \Pi\du i\ulb \Pi\du j{\ul\a} \Big[
- i \fracm 1 {\sqrt 2} (\s^\ulc \l^I ) _{\ul\a} F_{\ulb\ulc}
{}^I \Big] \bigg\}~~.
&(4.7)  \cr}  $$
We easily see that all of these $~\b\-$functions vanish
exactly after the use of the superfield equations (2.4) - (2.8).
Recall the definition of the $~\b\-$functions at the $~K\-$loop order:
$$\b = \sum_{L=1} ^K L T_{(1)}^{(L)}~~,
\eqno(4.8) $$
where $~T_{(1)}^{(L)}$~ is the coefficient of the $~1/\e\-$pole at the
$~L\-$loop order.  For example,
the graviton $~\b\-$function $~\b_{(\ula\ulb)}$~ is given by the
coefficient for the $~\Pi\du i\ula \Pi^{i\ulb}\-$term in $~\Lag_{\rm SG}
^{\infty (L=1)} + 2\Lag_{\rm SG}^{\infty(L=2)}$~ with the extra factor 2
for $~L=2$.  More explicitly we have
$$\b_{(\ula\ulb)} \, : ~~ \Pi\du i\ula \Pi^{i\ulb}
\left[ R_{(\ula\ulb)} + 4 \nabla_{(\ula} \nabla_{\ulb)} \Phi
- 4F\du \ula {\ulc I} F\du{\ulb\ulc} I
+ i\l^I \s_{(\ula}\nabla_{\ulb)} \Bar\l{} ^I + i\Bar\l{}^I
\s_{(\ula} \nabla_{\ulb)} \l^I \right] ~~,
\eqno(4.9) $$
and this vanishes exactly by the graviton superfield equation (2.4).

        As we have seen, all the computations in our superspace
for the $~D=(2,2)$~ space-time is just parallel to the ordinary one, and
there arises no further complications, as long as we use the {\it underlined}
indices for the vectorial indices.

\bigskip\bigskip

\noindent 5. {\it Self Duality Conditions on YM and SG Backgrounds.}~~~So
far our backgrounds are {\it not} subject to the SD condition, whose
consistency with our GS string is to be now studied.
As we have presented in our recent paper [8], the {\it supersymmetric}
SD conditions on the YM and SG backgrounds are the requirement of the
Majorana-Weyl spinor conditions [8] on the gaugino and the gravitino field
strength:
$$ \li{& \l\du \a I = 0 ~~,
&(5.1) \cr
& W_{\a\b\g} = 0~~,
&(5.2) \cr } $$
where $~W_{\a\b\g}$~ is the $~D=(2,2)$~ analog of the usual irreducible
SG superfield strength [10].
We first inspect the {\it purely} YM background, which is supposed
to correspond to the {\it open} GS string.
The spinorial condition (5.1) triggers the usual SD condition on the YM
field strength [8]:
$$F_{a b}{}^I = 0~~, ~~~~ F_{\Bar a\Bar b}{}^I = 0~~, ~~~~
\eta^{a \Bar b} F_{a\Bar b}{}^I = 0 ~~.
\eqno(5.3) $$
The consistency of (5.1) and (5.3) with all the YM BIds.~has
been already inspected in our recent paper [8].  In particular, we
recall the remarkable fact that the disappearance of the $~\l^2\-$type source
term on the r.h.s.~of the YM field equation (2.8) is consistent with
the SD condition (5.3).

        Our next important
question is the consistency of (5.1) with the $~\k\-$symmetry (3.3).
This can be easily answered by taking its $~\k\-$transformation under
(3.3):
$$\d_\k \,\l_\a{}^I  = (\d_\k E^{\ul\g} )
\left[ \nabla_{\ul\g} \l\du\a I - \phi\du{\ul\g\a}\d \l\du\d I
- f^{I J K} A\du {\ul\g} J \l\du\a K \right] ~~.
\eqno(5.4) $$
Obviously all the terms contains $~\l_\a{}^I$, so that
the original condition (5.1) is consistent with our
$~\k\-$transformation.  This consistency provokes
the nice idea of the constraint lagrangian to be added to the original
GS lagrangian (3.1) and (3.2):
$$I_{\rm C} \equiv \int d^2 \s \, \left[\,\Pi\du\plpl\a \l_\a{}^I
\,\right] i (\Bar\psi_-T^I\psi_-) \equiv\int d^2 \s\,
\Lag_{\rm C}~~.
\eqno(5.5)$$
We see that all the terms in the
$~\k\-$variation of $~\Lag_{\rm C}$~ are proportional to the condition
(5.1), and the $~\k\-$invariance of the total action $~I_{\rm SG} +
I_{\rm YM} + I_{\rm C} $~ is guaranteed.  As for the satisfaction of the
$~\b\-$functions, they are automatically satisfied, due to the fact that
the {\it supersymmetric} SDYM system is the subset of the solutions of
the general superfield equations (2.4) - (2.8).

We have so far ignored the curved SG background, which is more
subtle.  Let us first look into the {\it free} SG superfield
equations in (2.4) and (2.6) even {\it without} the tensor multiplet.
Recall that the SD condition for the Riemann tensor is to be
$$ R \du {a b}{\ulc\uld}  = 0 ~~, ~~~~
\eta^{a \Bar b} R\du{a\Bar      b}{\ulc\uld} = 0~~.
\eqno(5.6) $$
For a K{\" a}hler manifold {\it without} torsion, these conditions for SD are
equivalent to the Ricci-flatness condition $~R_{\ula \ulb} = 0$.
When higher-order interactions
in $~\a{\hskip 0.02pt}{}'$~ or other multiplets are included,
the condition (5.6) is violated unfortunately.  For example, the
$~F^2\-$term in the r.h.s.~of (2.4) does {\it not} vanish, and this
spoils the original condition $~R_{\ula\ulb} = 0$.
This feature has been known in the context of the Calabi-Yau compactification
for $~N=(1,0)$~ superstring models [16].  The conditions like $~\eta^{a\Bar b}
R_{a\Bar b}{}^{\ulc\uld} = 0$~ is {\it modified} in the presence
of higher-loop corrections
in the $~\s\-$model, but they can be eventually recovered by means of
{\it non-local} field redefinitions [16].  This has first popped up
as the breaking of the Ricci-flatness condition in $~N=(1,0)~$
superstring [16].  It seems to be a common fate of the superstring theories
that such initial condition specifying the manifolds
is lost at higher-order.  The same seems also true for our
(super) SDYM condition (5.1) and (5.5).  Actually, in a recent
paper using the vertex operators in the light-cone gauge [3], the
corresponding modifications for the YM and SG are reported.  Since the
analysis including the SG sector will be messy, we do not give their details
in this paper.

In our treatment of SG as well as of YM we did not include auxiliary
fields,\footnotew{Neeless to say, the content of auxiliary fields is
supposed to coincide with that of $~D=(1,3)$~ case [10].} which might
have helped in such context as getting the field equations on the
{\it flat} SG background in
(2.7) and (2.8).  However, it turned out that they do not provide
advantages for the SD condition.  This is because
the SDYM vector multiplet
can {\it not} be formulated {\it off-shell}, even though auxiliary
fields are provided [8].  The SDYM or SDSG multiplet is essentially
{\it on-shell} supersymmetric system.

\bigskip\bigskip

\noindent 7. {\it Concluding Remarks.}~~~In this Letter we have presented
a superspace formulation of $~D=(2,2),\,N=1$~ supersymmetric YM multiplet
coupled to SG multiplet in the space-time with the indefinite signature
$~(+,+,-,-)$.  We next formulated a GS
$~\s\-$model, and inspected its $~\k\-$invariance.  We derived the
$~\b\-$functions for those backgrounds, and saw that the condition of vanishing
$~\b\-$functions are consistent with the superfield equations of YM plus
SG up to the three-loop for YM and up to the two-loop for SG.  We also
inspected the validity of the {\it supersymmetric} SDYM condition with the
GS action, which is dictated by (5.1).  This result
shows that the {\it supersymmetric} SDYM theory is really
the consistent background
for the GS string, namely the consistent {\it low energy effective theory} of
$~N=(2,0)$~ heterotic GS superstring.

We inspected that the SD
condition can be imposed on these backgrounds
consistently with the GS $~\s\-$model.  Our formulation may provide
the foundation for many different future studies, such as the exhaustive
investigations of {\it all} possible dimensional reductions/compactifications
[1] of our starting super SDYM and SDSG in space-time with the appropriate
signature $~(+,+,-,-)$, which is supposed to generate {\it all}
possible exactly soluble supersymmetric systems in lower-dimensions.

        To our knowledge, the GS superstring in the non-compact
space-time with the signature $~(+,+,-,-)$~ has {\it never} been explicitly
presented in the past, not to mention the couplings to
the {\it supersymmetric} SDYM and SDSG multiplets.  Our GS formulation
is the {\it first} explicit and interesting trial,
motivated by $~N=2$~ superstring in the {\it supersymmetric} SDYM
or SDSG background.  In particular, the {\it
supersymmetric} SD condition based on the Majorana-Weyl spinors [8],
indicates yet unknown interesting features of
the superstring system worthwhile to study into the details [15].

There have been in the past several attempts to construct invariant
lagrangians in terms of {\it covariant} quantities.  However, it may
well be that such a lagrangian does {\it not} exist in $~D=(2,2)$, but it
makes sense only in terms of path-integral in the string action.
Also from this viewpoint, the construction of GS superstring has its own
important significance for the {\it supersymmetric} SDYM or SDSG.

        As has been already mentioned, we can conjecture that our
GS superstring theory will generate {\it all} the {\it exactly soluble
supersymmetric systems} in lower-dimensions.  Amusingly, while the usual
$~N=(1,0)~$ superstring has been suggested as the ``theory of everything'',
our GS superstring seems to be the ``theory of every soluble
supersymmetric thing''.  Remarkably, such simple systems as the
supersymmetric soluble theories themselves have a more fundamental theory
underlying them!  It is also interesting to realize that the
two-dimensional superconformal
theories as the bases of superstring theories themselves are also
soluble supersymmetric systems, generated by our GS superstring!
In this sense, we can develop a viewpoint that our GS
superstring is the {\it most} fundamental underlying superstring
theory, from which all other superstring theories originate.

\bigskip\bigskip

\noindent {\bf ACKNOWLEDGEMENT}

We appreciate D.~Depireux for his participation in an earlier stage of
this work.

\vfill\eject

\baselineskip 6 pt
\refs

{}~~~

\items{1} A.A.~Belavin, A.M.~Polyakov, A.~Schwarz and Y.~Tyupkin,
\pl{59}{75}{85};
\item{ } R.S.~Ward, \pl{61}{77}{81};
\item{ } M.F.~Atiyah and R.S.~Ward, \cmp{55}{77}{117};
\item{ } E.F.~Corrigan, D.B.~Fairlie, R.C.~Yates and P.~Goddard,
\cmp{58}{78}{223};
\item{ } E.~Witten, \prl{38}{77}{121};
\item{ } A.N.~Leznov and M.V.~Saveliev, \cmp{74}{80}{111};
\item{ } L.~Mason and G.~Sparling, \pl{137A}{89}{29};
\item{ } I.~Bakas and D.A.~Depireux, \mpl{6}{91}{399};
Univ.~of Maryland preprint, UMDEPP-91-111;
\mpl{6}{91}{1561}, Erratum-ibid {\bf {A6}} (1991) 2351.

\items{2} M.F.~Atiyah, unpublished;
\item{  } R.S.~Ward, Phil.~Trans.~Roy.~Lond.~{\bf A315} (1985) 451;
\item{  } N.~Hitchin, Proc.~Lond.~Math.~Soc.~{\bf 55} (1987) 59;
\item{  } A.A.~Belavin and V.E.~Zakharov, \pl{73}{78}{53}.

\items{3} H.~Ooguri and C.~Vafa, \mpl{5}{90}{1389}; \np{361}{91}{469};
\item{  } Kyoto RIMS preprint, RIMS-766 (July 1991).

\items{4} M.~Ademollo, L.~Brink, A.~D'Adda,
R.~D'Auria, E.~Napolitano, S.~Sciuto, E.~Del Giudice, P.~Di Vecchia,
S.~Ferrara, F.~Gliozzi, R.~Musto, R.~Pettorino, J.~Schwarz,
\np{111}{76}{77}.

\items{5} H.~Nishino and S.J.~Gates, Jr., Maryland preprint, UMDEPP-92-137
(Jan.~1992).

\items{6}  See e.g., M.B.~Green, J.H.~Schwarz and E. Witten, {\it Superstring
Theory},
\item{  }  Vols.~I and II, Cambridge University Press (1987).

\items{7} E.~Bergshoeff, H.~Nishino and E.~Sezgin, \pl{166}{86}{141}.

\items{8}  S.V.~Ketov, S.J.~Gates, Jr.~and H.~Nishino, Univ.~of Maryland
preprint, UMDEPP-92-163 (Feb.~1992).

\items{9} I.~Volovich, \pl{123}{83}{329};
\item{  } C.R.~Gilson, I.~Martin, A.~Restuccia and J.G.~Taylor,
\cmp{107}{86}{377}.

\items{10} S.J.~Gates, Jr., M.T.~Grisaru, M.~Ro{\v c}ek and W.~Siegel,
``{\it Superspace}'', (Benjamin/Cummings, Reading MA, 1983).

\items{11} A.~Salam and E.~Sezgin, {\it ``Supergravities in Diverse
Dimensions''}, Elsevier Science
Publishings, B.V.~and World Scientific Pub.~Co.~Pte.~Ltd.~(1989).

\items{12} H.~Nishino, unpublished (1988).

\items{13} M.T.~Grisaru, H.~Nishino and D.~Zanon,
\pl{206}{88}{625};
\item{  }  \np{314}{89}{363}.

\items{14} H.~Nishino, \np{338}{90}{386}.

\items{15} S.J.~Gates. Jr., S.V.~Ketov and H.~Nishino, Maryland preprint,
in preparation (March 1992).

\items{16} E.~Witten, \np{268}{86}{79};
\item{  } A.~Sen, \np{228}{86}{287};
\item{  } C.M.~Hull, \pl{178}{86}{357};
\item{  } M.T.~Grisaru, Van de Ven and D.~Zanon, \pl{173}{86}{423};
\item{  } \np{277}{86}{388}.

\vfill\eject
\end{document}